\documentstyle[sprocl]{article}

\input{psfig}
\def\be{\begin{equation}}
\def\ee{\end{equation}}
\def\bea{\begin{eqnarray}}
\def\eea{\end{eqnarray}}

\begin{document}

\title{ROUGHENING OF  ION-ERODED SURFACES}

\author{A.-L. BARAB\'ASI, M. A. MAKEEV, C. S. LEE }

\address{Department of Physics, University of Notre Dame, Notre
Dame,
IN 46556}

\author{R. CUERNO}

\address{Departmento de Matem\'aticas and Grupo Interdisciplinar de 
Sistemas Complicados, Universidad Carlos III de Madrid, 
c/ Butarque 15, 28911 Legan\'es, SPAIN}


\maketitle\abstracts{Recent experimental studies focusing on the
morphological properties of surfaces eroded by ion-bombardment report
the observation of self-affine fractal surfaces, while others provide
evidence about the development of a  periodic ripple
structure. To explain these discrepancies we derive a stochastic
growth equation that describes the evolution of surfaces eroded by ion
bombardment.  The coefficients appearing in the equation can be
calculated explicitly in terms of the physical parameters
characterizing the sputtering process.  Exploring the connection
between the ion-sputtering problem and the Kardar-Parisi-Zhang and
Kuramoto-Sivashinsky equations, we find that morphological transitions
may take place when experimental parameters, such as the angle of
incidence of the incoming ions or their average penetration depth, are
varied. Furthermore, the discussed methods allow us to calculate
analytically the ion-induced surface diffusion coefficient, that can
be compared with experiments.  Finally, we use numerical simulations
of a one dimensional sputtering model to investigate certain aspects
of the ripple formation and roughening.}


\section{Introduction}

In the last decade we have witnessed the development of an array of
theoretical tools, ideas and techniques intended to describe and
characterize the growth and roughening of nonequilibrium surfaces
\cite{alb,revrough1,revrough2,revrough5}. Initiated
by advances in understanding the statistical mechanics of various
nonequilibrium systems, it has been observed that for most surfaces in
nature the roughness follows simple scaling laws.  These surfaces are
self-affine fractals, being characterized by the roughness or
self-affine exponent $\alpha$.  One of the main advantage of this description
is that various growth processes can be classified into universality
classes that share the same scaling exponents.  On the practical side
this means that the scaling exponents characterizing roughness do not
vary continuously, but are defined by the universality class to which
they belong.

     One particularly important thin film processing technique is ion
beam sputtering \cite{revsput,rev-sput,rev-sput2}. Sputtering is the
removal of material from the surface of solids through the impact of
energetic particles.  It is a widespread technique, used in a large
number of applications, with a remarkable level of sophistication. It
is a basic tool in surface analysis, depth profiling, sputter
cleaning, micromachining, and sputter deposition.

  Motivated by the advances in understanding growth, and by the
need
of having a detailed knowledge on the morphology of the sputter
eroded
surfaces, recently a number of experimental studies have
investigated
the morphological properties of surfaces eroded by ion
bombardment.
Briefly, the experimental results can be classified in two main
classes. There exists ample evidence about the development of a
periodic ripple structure in sputter etched surfaces
\cite{carter,chason,Chason94b,Chason92,Chason93,maclaren}.
However,
a
number of recent investigations have provided rather detailed and
convincing experimental evidence, that under certain experimental
conditions ion-eroded surfaces are rough and self-affine, and the
roughness follows the predictions of various scaling theories
\cite{eklund,krim,yang}. Moreover, these investigations
did
not find
evidence of ripple formation on the surface!

The discrepancy between the results of the mentioned investigations
motivated us to have a second look at the mechanisms shaping the
morphology of ion eroded surfaces \cite{us}.  In this paper we
investigate the large scale properties of ion-sputtered surfaces
aiming to understand in an unified framework the various dynamic and
scaling behaviors of the experimentally observed surfaces.  For this
we derive a stochastic nonlinear equation that describes the time
evolution of the surface height.  The coefficients appearing in the
equation are functions of the physical parameters characterizing the
sputtering process.  We find that transitions may take place between
various surface morphologies as the experimental parameters (e.g.\
angle of incidence, penetration depth) are varied.  Namely, at short
length-scales the equation describes the development of a periodic
ripple structure, while at larger length-scales the surface morphology
may be either logarithmically ($\alpha=0$) or algebraically ($\alpha >
0$) rough.  Furthermore, we calculate analytically the ion-induced
diffusion constant, $D^I$, and its dependence on the ion energy, flux,
angle of incidence, and penetration depth.  We find that there exists
a parameter range when ion bombardment generates a {\it negative}
surface diffusion constant, leading to morphological instabilities
along the surface, affecting the surface roughness and the ripple
structure.  The effect of ion-induced diffusion on the morphology of
ion-sputtered surfaces is summarized in a morphological phase diagram,
allowing for direct experimental verification of our predictions.
Finally, we use numerical simulations
of a one dimensional sputtering model to investigate certain aspects
of the ripple formation and roughening.

\section{Scaling theory}

A common feature
of most non-equilibrium rough interfaces \cite{alb,revrough1,revrough2}
observed experimentally or in discrete models
is that their roughening follows simple scaling laws.  The associated
scaling exponents can be obtained using numerical simulations or
stochastic evolution equations.  The morphology and dynamics of a
two-dimensional rough surface can be characterized with the 
{\it interface width\/},
defined by the rms fluctuation in the height variable $h(x,y,t)$,
\begin{equation}
\label{width-def}
W(L,t)\equiv \sqrt{{1 \over {L^2}} \langle \sum_{x,y=1,L}
[h(x,y,t)-\bar
h]^2} \rangle,
\end{equation}
where $L$ is the linear size of the sample,  the brackets $\langle...\rangle$
denote  ensemble  average, and
the {\it mean height\/} of the surface, $\bar h$, is defined by
\begin{equation}
\overline{h(t)} \equiv {1 \over {L^2}} \sum_{x,y=1,L} h(x,y,t).
\label{mean-def}
\end{equation}
For times $t \gg t_{\times} \sim L^z$, the surface  width behaves as
\begin{equation}
W(L,t) \sim L^\alpha,
\label{scal}
\end{equation}
where $\alpha$ is the {\it roughness exponent} and $z$ is the {\em dynamic
exponent}. Regarding the early dynamics
of the roughening process, the total width increases as
$W(L,t) \sim t^\beta$, where $\beta$ is the {\it growth
exponent}. The dynamic exponent is related to
$\alpha$ and $\beta$ as $z=\alpha/\beta$ \cite{Family85}.

To understand the roughening process,  we need to develop
methods
to
predict the value of the scaling exponents $\alpha$ and $\beta$. 
A
breakthrough in this direction was the introduction of the
Kardar-Parisi-Zhang (KPZ) equation \cite{kpz}
\begin{equation}
\frac{\partial h}{\partial t} =  \nu \nabla^2 h + \lambda (\nabla
h)^2 +
\eta({ x, y},t)~~~~~[KPZ].
\label{e:kpz} 
\end{equation}
The first term on the rhs describes the relaxation of the
interface
due to the surface tension $\nu$ and the second is a generic
nonlinear
term incorporating lateral growth. The noise, $\eta({ x},y,t)$,
reflects the random fluctuations in the growth process and is an
uncorrelated random number that has zero configurational average. 
For
one dimensional interfaces the scaling exponents of the KPZ
equation
are known exactly, as $\alpha=1/2$, $\beta=1/3$, and $z=3/2$.
However,
for higher dimensions they are known only from numerical
simulations.
For the physically most relevant two dimensional interface we
have
$\alpha \simeq 0.38$ and $\beta \simeq 0.18$
\cite{numer}.

If $\lambda=0$ in (\ref{e:kpz}), the remaining equation describes
the equilibrium fluctuations of an interface which tries to minimize 
its area. This equation, introduced
and studied in the context of interface roughening by Edwards and
Wilkinson (EW) \cite{Edwards82}, can be solved exactly due to its
linear character, giving the scaling exponents $\alpha=(2-d)/2$
and
$\beta=(2-d)/4$. For two dimensional interfaces we have
$\alpha=\beta=0$, leading to a logarithmic roughening of the
interface.

\section{Experimental results}

The morphology of surfaces bombarded by energetic ions has long
fascinated the experimental community.  Lately, with the development
of high resolution observation techniques, this question is living a
new life.

We shall focus here on two dominant morphologies, ripple formation and
kinetic roughening, since these are observed in the sputtering of
impurity free, amorphous materials. Impurities that bind strongly to
the surface (being thus difficult to sputter) may induce
other dominant morphological features, such as cones or abrupt walls
\cite{carter}. These will not be considered in this paper. Also, we
limit ourselves to sputtering by ion bombardment, in which the ions
have parallel trajectories and the same velocity. Thus we will not
consider plasma etching (where the ions have a broad energy
distribution and random angles of incidence) or chemical sputtering,
where the yield is influenced by the chemical reactions taking place
on the surface.

\subsection{Ripple  formation}

Ripple formation on ion-sputtered surfaces have been observed by
many
groups in various systems and ion beams (for a review see
\cite{carter}). Here  we discuss  a few recent investigations
that
characterized in great detail the observed morphologies.

Evidence for the ripple structure on the surfaces of SiO$_2$ and Ge
has been provided in a series of studies by Chason {\it et al.\/}
\cite{chason,Chason94b,Chason92,Chason93}. We shall discuss here the
results obtained on SiO$_2$ \cite{Chason94b,Chason92}. A low energy
ion beam (Xe, H or He), with energies $\le$ 1 keV is directed towards
a SiO$_2$ sample with an angle of incidence of 55$^\circ$ from normal.
The typical incoming flux is 10$^{13}$ cm$^{-2}$s$^{-1}$. The
interfaces are analyzed using {\it in situ\/} energy dispersive x-ray
reflectivity and {\it ex situ} atomic force microscopy (AFM).
Bombarding the surface with 1 keV Xe ions, one finds that the
interface roughness, determined from X-ray diffraction, increases
linearly with the fluence (the fluence is the number of incoming atoms
per surface area, and plays the role of time in these
measurements). Thus $\beta=1$, too large a value to be interpretable
by continuum theories. Such a large value of $\beta$ indicates the
existence of an instability in the system. Physically, the instability is
balanced by surface diffusion, leading to the appearance of
the ripple structure whose wavelength increases with temperature.  
Such a ripple structure can be seen if one
inspects the AFM pictures of the interface. A similar ripple structure
has been observed for Ge surfaces bombarded by Xe atoms \cite{chason}.

Another series of experiments on ripple formation were reported by
MacLaren {\it et al.} \cite{maclaren}. They studied InP and GaAs
bombardment with 5 keV Ar$^+$, 17 keV Cs$^+$ and 5.5 keV O$_2^+$ beams
in a temperature range of $-$50 to 200 $^\circ$C.  Their study
revealed in detail the temperature dependence of the ripple
wavelength. For example for GaAs bombarded by Cs$^+$ ions the ripple
spacing increased from zero 0.89 $\mu$m to 2.0 $\mu$m, as the
temperature increased from 0 $^\circ$C to 100 $^\circ$C. Probably the
most interesting finding of their study was that when lowering the
temperature, the ripple spacing (wavelength) did not go continuously
to zero, as one would expect, since the diffusion constant decreases
exponentially with the inverse temperature, but rather at around 20
$^\circ$C it stabilized at an approximately constant value.  MacLaren
{\it et al.} interpreted this as the emergence of a radiation enhanced
diffusion, that gives a constant (temperature independent)
contribution to the diffusion constant. We shall return to ion
enhanced diffusion in Sects. 6. and 7.

\subsection{Roughening}

For graphite bombarded with 5 keV Ar ions, Eklund  {\em et al.}  
\cite{eklund} reported
$\alpha \simeq
0.2-0.4$, and $z \simeq 1.6-1.8$, values consistent with the
predictions of the Kardar-Parisi-Zhang (KPZ) equation in 2+1
dimensions \cite{kpz,numer}. No trace was found of a periodic ripple 
structure.  In these  experiments 
pyrolytic
graphite was bombarded by 5 keV Ar ions, which arrived with an
angle
of incidence of 60$^\circ$. The experiments were carried out for
two
flux values, $6.9 \times 10^{13}$ and $3.5 \times 10^{14}$
ions/cm$^2$, and the total fluences obtained were $10^{16}$,
$10^{17}$
and $10^{18}$.  The etched graphite was examined using STM.  
Large scale features develop with continuous bombardment,
the interface becoming highly correlated and rough. 

A somewhat larger roughness exponent has been measured for
samples of
iron bombarded with 5 keV Ar arriving with angle of incidence of
25$^\circ$. The interface morphology was observed using STM, and
the height--height correlation function results in a roughness
exponent
$\alpha = 0.53 \pm 0.02$ \cite{krim}. The mechanism leading to
such
a
roughness exponent is not yet understood in terms of the
continuum
theories, since for two dimensions the growth equations predict
0.38,
2/3 and 1, all values far from the observed value.

Finally Si(111) sputtered by 0.5 keV Ar$^+$ ions has also been observed to
 roughen, in this case following an anomalous dynamic scaling form $w(t,\ell)
\sim
 ln(t) \ell^{2 \alpha}$, with $\alpha \simeq 1.15 \pm 0.08$
 \cite{yang}, where $w(t,\ell) $ is the surface width for a small 
 window of lateral size $\ell$.

\section{Continuum theory}

\subsection{Sigmund's theory of sputtering}

In order to calculate the sputtering yield, and predict the
surface morphology, we first need to understand   the mechanism of
sputtering, resulting from the interaction of the incident ion
and the surface layer. 

\begin{figure}[bht]
\hskip 3.0 cm
\psfig{figure=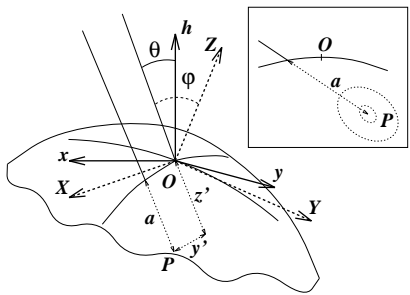,height=1.5in}
\caption{Reference frames for the computation of the erosion 
velocity at point $O$. Inset: Following a straight trajectory
(solid line) the ion penetrates an average distance $a$ inside
the solid (dotted line) after which it completely spreads out
its kinetic energy. The dotted curves are equal energy contours.
Energy released at point $P$ contributes to erosion at $O$.}
\label{fig1}
\end{figure}

A qualitative picture is as follows\cite{carter} (see
Fig. \ref{fig1}).  The incoming ions penetrate the surface and
transfer their kinetic energy to the atoms of the substrate by
colliding with the substrate atoms, or through other processes such as
electronic excitations. Atoms that recoil with sufficient energy
undergo secondary collisions, thereby generating another generation of
recoiling atoms.  A vast majority of atoms will not gain enough energy
to leave their lattice position permanently.  However, some are
permanently removed from their sites, locally making the substrate
amorphous. The atoms that are near the surface and gain enough energy
to break their bonds and leave the surface will be sputtered. The
scattering events that might lead to sputtering take place within a
certain layer of average depth $a$. Usually the number of sputtered
atoms is orders of magnitudes smaller than the total number of atoms
participating in the collision cascade.

A rather successful theory of the above processes was introduced by 
Sigmund to describe the experimentally observed sputtering yields 
\cite{sig}. His treatment considers the energy transfer from the incoming 
ion to the atoms of an
isotropic solid by writing down a Boltzmann transport equation for the
atoms. Expanding this equation in form of Legendre polynomials, he
obtains a solution using the method of moments.
One of the most important result of his analysis is that   for
low
energies the damage and energy distribution generated by the
incoming  ion follows a Gaussian.
Thus here, following \cite{sig,bh}, we consider that the
average
energy
deposited at point $O$ due to the ion arriving at $P$ follows the
Gaussian distribution
\begin{equation}
E(\mbox{\boldmath $r'$})  =  \frac{\epsilon}{(2\pi)^{3/2} \sigma
\mu^2} \exp \left\{ - \frac{z'^2}{2 \sigma^2} -
\frac{x'^2+y'^2}{2 \mu^2} \right\}.
\label{gaussian}
\end{equation}
In (\ref{gaussian}) $z'$ is the distance measured along the ion
trajectory, and $x'$, $y'$ are measured in the plane
perpendicular
to
it (see Fig.\ \ref{fig1}; for simplicity in the figure $x'$ has
been
set to 0); $\epsilon$ denotes the total energy carried by the ion
and
$\sigma$ and $\mu$ are the widths of the distribution in
directions
parallel and perpendicular to the incoming beam,  respectively.
However, the sample is subject to an uniform flux $J$ of
bombarding
ions. A large number of ions penetrate the solid at different
points
simultaneously and the velocity of erosion at $O$ depends on the
total
power ${\cal E}_O$ contributed by all the ions deposited within
the
range of the distribution (\ref{gaussian}).  If we ignore
shadowing
effects among neighboring points, as well as further redeposition
of
the eroded material, the normal velocity of erosion at $O$ is
given
by
\begin{equation}
v = p \; \int_{{\cal R}} d\mbox{\boldmath $r$} \;
\Phi(\mbox{\boldmath $r$})
\; E(\mbox{\boldmath $r$}),
\label{vel}
\end{equation}
where the integral is taken over the region ${\cal R}$ of all the
points
at which the deposited energy contributes to ${\cal E}_O$,
$\Phi(\mbox{\boldmath $r$})$
is a local correction to the uniform flux $J$ and $p$ is a
proportionality
constant between power deposition and rate of erosion.
In the following we review the basic steps in the calculation
of $v$; further details can be found in Refs.\ \cite{us,cb,bh}.

\subsection{Continuum equation for the surface height}

In this section  we derive an equation of motion for the surface
height from the physical model of ion--sputter erosion discussed
in
the previous section.  Since we are mainly interested in the
physically relevant case of a two dimensional substrate and the
one
dimensional case to linear order is very clearly explained in the
work
by Bradley and Harper \cite{bh}, we refer the reader to that
reference, and focus here on the more general 2d case. 

In the following we summarize the steps in the derivation of the
equation of
motion. 

{\bf (i)} First we calculate the normal component of the velocity
of
erosion $v_O$ at a generic point $O$ of the interface. This
calculation is most easily performed in a local frame of
reference
$(X,Y,Z)$ defined as follows: the $\hat{Z}$ axis is identified
with
the normal direction to the average surface orientation at $O$. 
Moreover $\hat{Z}$ forms a plane with the trajectory of an ion penetrating
the
surface at $O$. We choose the $\hat{X}$ axis to lie in that
plane. Finally, $\hat{Y}$ is the remaining direction which
completes a right--handed reference frame, see Fig.\ \ref{fig1}.

{\bf (ii)} Next we relate the quantities measured in coordinates
of
the local frame to coordinates in the laboratory frame $(x,y,h)$. 
The
latter is defined by the experimental configuration. That is, $h$
is
the direction normal to the uneroded flat surface. The ion
trajectories together with the $h$ axis define a plane, which is
taken
to be the $x-h$ plane. And finally the $y$ axis completes a
right--handed reference frame, see Fig.\ \ref{fig1}.  However,
$\varphi$, which is the angle between the ion trajectory and the
{\it
local} normal to the surface, changes from point to point along
the
surface, and is a function of the local values of the slopes at
$O$
(as seen in the laboratory frame), as well as of the fixed angle
$\theta$ subtended by the ion trajectories and the normal to the
uneroded surface (the $h$ direction in Fig.\ \ref{fig1}).

{\bf (iii)} In the absence of overhangs the surface can be described by a
single valued height function $h(x,y,t)$, measured from an initial flat
configuration which is taken to lie in the ($x$,$y$) plane.  The
ion beam is parallel to the $x$-$h$ plane forming an angle $0 \leq 
\theta < \pi/2$ with the $z$ axis. To obtain the equation of motion for the
surface profile function $h(x,y,t)$, we will have to project the normal
component of the velocity of erosion onto the global $h$ axis. We get
\begin{equation}
\frac{\partial h(x,y,t)}{\partial t} \simeq -
v(\varphi,R_X,R_Y) \sqrt{1+ (\nabla h)^2} ,
\label{hv}
\end{equation}
where $\varphi$ is the angle of the beam direction with the local
normal to the surface at $h(x,y,t)$ and $R_{X,Y}$ the values
of the local radii of curvature at $(x,y,h)$. Now $\varphi$ is a function of
the angle of incidence $\theta$ and the values of the local slopes
$\partial_x h$ and $\partial_y h$, and can be expanded in powers
of the latter. We will assume that the surface varies smoothly
enough so that products of derivatives of $h$ can be neglected for third or
higher orders.

At this stage additional relevant physical processes must be
taken
into account to describe the evolution of the surface. First, the
bombarding ions reach the surface at random positions and times.
We
account for the stochastic arrival of ions by adding to
(\ref{hv})
a
Gaussian white noise $\eta(x,y,t)$ with zero mean and variance
proportional to the flux $J$.  Second, at finite temperature
atoms
diffuse on the surface \cite{chason,eklund}.  To include
this surface
self-diffusion we allow for a term $-D_T \nabla^2 (\nabla^2 h)$
\cite{herr,Villain91a}, where $D_T$ is a
temperature
dependent positive coefficient.  Expanding (\ref{hv}) and adding
the
noise and the surface-diffusion terms we obtain the equation of
motion
\cite{notexi}
\begin{eqnarray}
\frac{\partial h}{\partial t} & = & -v_0 + \gamma \frac{\partial
h}{\partial x} + \nu_x \frac{\partial^2 h}{\partial x^2} + \nu_y
\frac{\partial^2 h}{\partial y^2} + \frac{\lambda_x}{2} \left(
\frac{\partial h}{\partial x} \right)^2 + \frac{\lambda_y}{2} \left(
\frac{\partial h}{\partial y} \right)^2 \nonumber \\ && - D^I_x
\frac{\partial^4 h}{\partial x^4} - D^I_y \frac{\partial^4 h}{\partial
y^4} - D_T \nabla^2 (\nabla^2 h) + \eta .
\label{eqn}
\end{eqnarray}
 From (\ref{hv}) we can compute the expressions for the coefficients
appearing in (\ref{eqn}) in terms of the physical parameters
characterizing the sputtering process.  To simplify the discussion we
restrict ourselves to the symmetric case $\sigma=\mu$.  The general
case is discussed in \cite{us}.  If we write $F \equiv (\epsilon J
p/\sqrt{2\pi}) \exp(-a_{\sigma}^2/2 -a_{\sigma}^2 s^2)$, $s \equiv
\sin \theta$, $c \equiv \cos \theta$ and $a_{\sigma} \equiv a/\sigma$,
we find for the coefficients in (\ref{eqn})
\begin{eqnarray}
v_0 & = & \frac{F}{\sigma} c \;\;,\;\; \gamma = \frac{F}{\sigma}
s
(a_{\sigma}^2 c^2 - 1) , \nonumber \\
\lambda_x & = & \frac{F}{\sigma} c \left\{ a_{\sigma}^2
(3s^2-c^2) - a_{\sigma}^4 s^2 c^2 \right\} , \nonumber \\
\lambda_y & = & - \frac{F}{\sigma} c \{ a_{\sigma}^2 c^2 \},
\label{coefs2} \\
\nu_x & = & \frac{F}{2} a_{\sigma} \left\{ 2 s^2- c^2 -
a_{\sigma}^2 s^2 c^2 \right\} , \nonumber \\
\nu_y & = & -\frac{F}{2} a_{\sigma} c^2,   \nonumber \\
D^I_x & = & \frac{F a^{2}}{24 a_{\sigma} } \left\{a_{\sigma}^4 s^4 c^2 +
 a_{\sigma}^2 ( 6 c^2 s^2 -4 s^4 ) + 3 c^2 -12 s^2 \right\},\nonumber \\
D^I_y& = &\frac{F a^{2}}{24 a_{\sigma} } 3 c^2.\nonumber
\end{eqnarray}

\section{Analysis of the obtained growth equations}

Consistent with the direction of the bombarding beam and the
choice
of
coordinates, the terms in (\ref{eqn}) are symmetric under $y
\rightarrow -y$ but not under $x \rightarrow -x$, while for
$\theta
\rightarrow 0$ we get $\gamma = \xi_x = \xi_y = 0$, $\lambda_x =
\lambda_y$ and $\nu_x = \nu_y$.  The equation studied
in
Ref.\ \cite{bh} corresponds to the deterministic linear version
of
(\ref{eqn}), i.\ e.\ $\lambda_x=\lambda_y=\eta=0$.

If $\nu_x$ and $\nu_y$ are positive, the surface diffusion term
is
expected to contribute negligibly to the relevant surface
relaxation
mechanism when we probe the system at increasingly large length
scales. Scaling properties are then described by the anisotropic
KPZ
equation (AKPZ), which predicts two possible behaviors depending
on
the relative signs of the coefficients $\lambda_x$ and
$\lambda_y$
\cite{wolf,bruins}.  If $\lambda_x \lambda_y > 0$, then
$\alpha=0.38$
and $z=1.6$, the surface width $W(L,t)$ increases algebraically,
being
characterized by the exponents of the KPZ equation in 2+1
dimensions
\cite{numer}. For $\lambda_x
\lambda_y
< 0$, the nonlinear terms
$\lambda_x$ and $\lambda_y$ become irrelevant, and the width
grows
only logarithmically, i.e.  $\alpha=0$.

 In our case $\nu_x$ can change sign as $\theta$ and $a_{\sigma}$
are
varied, while $\nu_y$ is always negative.  The negative $\nu$
causes
an instability, whose origin is the faster erosion for the bottom
of a
trough than for the peak of a crest, as predicted by (\ref{vel}) \cite{sig,bh}
(see also Fig.\ 3 of Ref.\ \cite{bh}).  An instability due to a
negative
surface tension is also known to take place in the
Kuramoto--Sivashinsky (KS) equation \cite{ks}, which is the {\em
noiseless} and isotropic version of (\ref{eqn}). It has been
argued
for the KS equation that in 1+1 dimensions $\nu$ renormalizes to
a
positive value \cite{ks1+1}, and the large length scale behavior
is
described by the KPZ equation.  In 2+1 dimensions it is not
completely
settled whether the large distance behaviors of KS and KPZ fall
in
the
same universality class, different approaches leading to
conflicting
results \cite{ks2+1}.

 In contrast to the KS equation, Eq.\ (\ref{eqn}) is anisotropic,
and
explicitly contains a noise term.  The competition between
surface
tension and surface diffusion generates a characteristic length
scale
in the system, $\ell_c = \sqrt{D_T/|\nu|}$, where $\nu$ is the
largest
in absolute value of the negative surface tension coefficients.
Below
we discuss a possible scenario for the scaling behavior predicted
by
(\ref{eqn}) based primarily on the results available in the
literature
for some of its limits.  The complete scaling picture should be
provided by either a DRG analysis capable of coping with the
linear
instabilities present in the system, or a numerical integration
of
(\ref{eqn}).

\begin{figure}[bht]
\hskip 3.5 cm
\psfig{figure=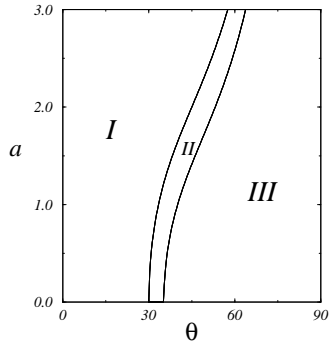,height=1.5in}
\vskip 0.5 cm
\caption{Phase diagram for the isotropic case $\sigma=\mu=1$. 
Region
I: $\nu_x<0$, $\nu_y <0$, $\lambda_x <0$, $\lambda_y < 0$; Region
II:
$\nu_x<0$, $\nu_y <0$, $\lambda_x >0$, $\lambda_y < 0$; Region III:
$\nu_x>0$, $\nu_y <0$, $\lambda_x >0$, $\lambda_y < 0$.  Here $a$
is
measured in arbitrary units and $\theta$ is measured in degrees.}
\label{fig2}
\end{figure}

The scaling behavior depends on the relative signs of $\nu_x$,
$\nu_y$, $\lambda_x$ and $\lambda_y$ \cite{note2}.  The
variations
of
these coefficients as functions of $a_{\sigma}$ and $\theta$ lead
to
the phase diagram shown in Fig.\ \ref{fig2}.

{\it Regions I and II---} For small $\theta$ both $\nu_x$ and
$\nu_y $
are negative. As discussed by Bradley and Harper \cite{bh} and
experimentally studied by Chason {\em et al.} \cite{chason}, a
periodic structure dominates the surface morphology, with ripples
oriented along the direction ($x$ or $y$) which presents the
largest
absolute value for its surface tension coefficient. The observed
wavelength of the ripples is $\lambda_c = 2 \pi \sqrt{2} \ell_c$.

The large length scale behavior $\ell \gg \ell_c$ is expected to
be
different. Now both nonlinearities and the noise may become
relevant.
The scaling properties of the surface morphologies predicted by
(\ref{eqn}) are unknown. A possible scenario is that the $\nu$'s
renormalize to positive values, as they do for the KS equation in
$1+1$ dimensions, and the large scale scaling properties of the
system
are described by the AKPZ equation.  Then one would observe
algebraic
scaling in region $I$, where both nonlinearities have the same
(negative) sign, whereas scaling would become logarithmic through
an
AKPZ-like mechanism in region $II$, where $\lambda_x$ and
$\lambda_y$
have opposite signs. Actually, the asymptotic KPZ scaling has been
recently shown to occur along the $\theta$ axis of Fig.\ 2
through a renormalization group analysis \cite{kent}.

{\it Region III ---} This region is characterized by a positive
$\nu_x$ and a negative $\nu_y$. Now the periodic structure
associated
with the instability is directed along the $y$ direction and is
the
dominant morphology at scales $\ell \sim \ell_c$. Again, such an
anisotropic and linearly unstable equation is unexplored in the
context of growth equations. Assuming that $\nu_y$ renormalizes
to
a
positive value, and that the AKPZ mechanism operates, one would
expect
logarithmic scaling in region $III$, since the nonlinear terms
have
opposite signs.

Even though several aspects of the scaling behavior predicted by
(\ref{eqn}) and (\ref{coefs2}) remain to be clarified, we believe that
these equations contain the relevant ingredients for understanding
roughening by ion bombardment \cite{last}.  To summarize, at short
length scales the morphology consists of a periodic structure oriented
along the direction determined by the largest in absolute value of the
negative surface tension coefficients \cite{chason}. Modifying the
values of $a_{\sigma}$ or $\theta$ changes the orientation of the
ripples \cite{carter,bh}.  At large length scales we expect two
different scaling regimes.  One is characterized by the KPZ exponents,
which {\it might} be observed in region $I$ in Fig.\ \ref{fig2}.
Indeed, the values of the exponents reported by Eklund {\em et al.}
\cite{eklund} are consistent within the experimental errors
with the KPZ exponents in 2+1 dimensions. The other regions ($II$ and
$III$) are characterized by logarithmic scaling ($\alpha=0$), which
has not been observed experimentally so far.  Moreover, by tuning the
values of $\theta$ and/or $a_{\sigma}$ one may induce transitions
among the different scaling behaviors.  For example, fixing
$a_{\sigma}$ and increasing the value of $\theta$ would lead from KPZ
scaling (region $I$) to logarithmic scaling ($II$, $III$) for large
enough angles.

Recent results by Rost and Krug on the two dimensional anisotropic KS
equation indicate that the scaling regimes II and III in the {\it
noiseless} limit of our model is dominated by exponentially growing
solutions of the KS equation \cite{rost}. In those regions the ripple
structure is oriented along a direction which is neither $x$ nor
$y$. Further numerical simulations are needed to understand the effect
of the noise on the stability of the exponential solutions.  Insight
into the expected morphologies is obtained from numerical simulation
of discrete models, correctly capturing the basic mechanisms taking
place during sputtering. Recent simulations on discrete models
indicate that the noisy KS equation indeed describes the dynamics of
the sputtering generated roughening \cite{rodo}.

\section{Ion-Induced Surface Diffusion in Ion Sputtering}

In the absence of ion bombardment surface diffusion is thermally
activated, and characterized by the diffusion constant, $D_T = D_0
\exp\left[- {E_d / k_B T}\right]$, such that the evolution of the
surface height, $h(x,y,t)$ is described by the continuum equation
${\partial h / \partial t} = - D_T \nabla^4 h$ \cite{herr}. Here $E_d$
is the activation energy for surface diffusion of the adatoms and $T$
is the substrate temperature.  However, numerous experiments regarding
the effect of ion bombardment on island formation, surface migration,
surface smoothing and ripple formation have provided evidence that ion
bombardment is accompanied by an increase in surface diffusion
\cite{maclaren,r2,r1,r3,r4,rossn}.  In particular, it has been
demonstrated that ion-induced surface diffusion can decrease the
epitaxial temperature \cite{r2}, enhance nucleation during growth
\cite{r1}, modify the surface morphology, or induce the existence of
a temperature independent ion-induced surface diffusion constant,
as in the experiments of MacLaren {\em et al.} referred to above.

\begin{figure}[bht]
\vskip 1.0 cm
\hskip 3.5 cm
\psfig{figure=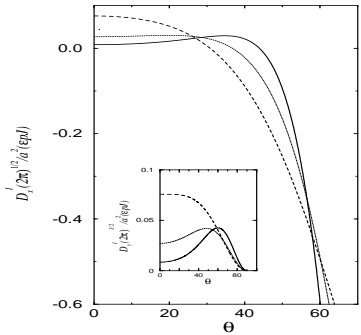,height=1.8in,width=2.0in}
\caption{Ion-induced diffusion constant, $D_x^I$ and $D_y^I$ (inset)
as a function of the angle of incidence $\theta$. In both figures the
curves correspond to $a_\sigma=1.5$ (dashed line), $a_\sigma=2.0$
(dotted line) and $a_\sigma=2.5$ (continuous line).}
\label{fig3}
\end{figure}

Although the effect of the ions on surface diffusion is well
documented experimentally, there is no theory that would quantify it.
Eq. (\ref{coefs2}) provides analytically the ion-induced diffusion
constant, $D^I$, and its dependence on the ion energy, flux, angle of
incidence, and penetration depth.
Consistent with symmetry considerations for $\theta=0$ we obtain
$D^I_x=D^I_y$. However, for $\theta \neq 0$ we find that $D^I_x \neq
D^I_y$, i.e. the ion-induced surface diffusion is {\it anisotropic}.
Moreover, its sign also depends on the experimental parameters.  Their
properties can be summarized as follows: (a) Independent of the angle
of incidence $D^I_y$ is positive, and decreases with $\theta$, while
the sign of the $D^I_x$ depends on both $\theta$ and $a_\sigma$ as
shown in Fig. \ref{fig3}.  Thus, while for $\theta=0$ the ion
bombardment enhances the surface diffusion ($D^I_{x} > 0$), for large
$\theta$ it can suppress diffusion; (b) The fact that $D_x^I$ can be
negative indicates that any simple theory connecting the magnitude of
the ion-induced diffusion to the energy transferred by the ions to the
surface is incomplete, since it can predict only a positive $D^I$. In
fact, $D^I$ is the result of a complex interplay between the local
{\it surface topography} and the energy transferred to the surface; (c)
The diffusion constants are proportional to the flux $J$, in agreement
with the detailed experimental study of Cavaille and Dreschner
\cite{r4}; (d) It is a standard experimental practice to report the
magnitude of the ion-enhanced diffusion using an effective temperature
$T^{eff}$ at which the substrate needs to be heated to obtain the same
mobility as with ion bombardment \cite{r4,rossn}.  We can calculate
$T^{eff}$ using the relation $D^I+ D_0 \exp(-E_a/k_BT) = D_0 \exp(-E_a
/k_B T^{eff})$, that has two important consequences.  First, the
anisotropic diffusion constant translates into an anisotropic
$T^{eff}$, i.e. we have $T^{eff}_x \neq T^{eff}_y$. The experimental
methods used to estimate $T^{eff}$ could not distinguish $T^{eff}_x$
and $T^{eff}_y$ \cite{r4,rossn}. However, current observational
methods should be able to detect the difference between the two
directions.  Second, while it is generally believed that ion
bombardment can only raise the effective temperature since it
transfers energy to the surface, the negative $D^I$ indicates that
along the $x$ direction one could have $T_{eff} < T$. (e) Finally,
the results (\ref{coefs2})  are based on 
Sigmund's theory of sputtering \cite{sig}, that
describes sputtering in the linear cascade regime. The energy
range when this approach is applicable lies between 0.5 keV and 1 MeV,
the precise lower and upper limits being material dependent.  Thus,  we
do not expect (\ref{coefs2}) to apply to low energy (few eV)
ion-enhanced epitaxy.

{\it Quantitative comparison with experiments---} At nonzero
temperature the total diffusion constant is given by $D= D^I +
D_T$. As $T$ decreases there is a critical temperature, $T_c$, at which
$D^I = D_T$, so that for $T<T_c$ the diffusion is dominated by its
ion-induced component, which is independent of temperature, in
agreement with the experimental results of MacLaren {\it et al.}
\cite{maclaren}.  Unfortunately, for most materials the quantities
entering in $F$, $a_\sigma$ and $D_0$ are either unknown, or only
their order of magnitude can be estimated. However, we can express
$T_c$ in terms of measurable quantities independent of these constants
\begin{equation}
T_c={T_0 \over 1-(2 T_0 k_B/E_a) \ln (\ell_{Ion}/\ell_{T_0})}
\label{tc}
\end{equation}
where $\ell_{T_0}$ is the experimentally measured ripple wavelength at
any temperature $T_0> T_c$; $\ell_{Ion}$ is the ripple wavelength in
the low temperature regime, $T < T_c$, where ion induced diffusion
dominates, and therefore $\ell_{Ion}$ is independent of $T$; $E_a$ is
provided by the slope of $\ln(\ell)$ versus $1/T$ in the high
temperature regime ($T >> T_c$). Consequently, {\it all} quantities in
(\ref{tc}) can be obtained from a plot of the ripple wavelength as a
function of temperature, so (\ref{tc}) gives $T_c$ {\it in terms of
measurable quantities}. Such a plot is provided by MacLaren {\it et al.}
\cite{maclaren}, leading to $E_a = 0.51$eV, $\ell_{Ion}=0.8\mu$m.
Using $\ell_{T_0}=2\mu$m for $T=368$K, we obtain $T_c=57^\circ$ C, which is
in good agreement with the experiments, that provide $T_c$ between 45
and 60$^\circ$ C \cite{maclaren}.

\begin{figure}[bht]
\vskip 0.5 cm
\hskip 3.5 cm
\psfig{figure=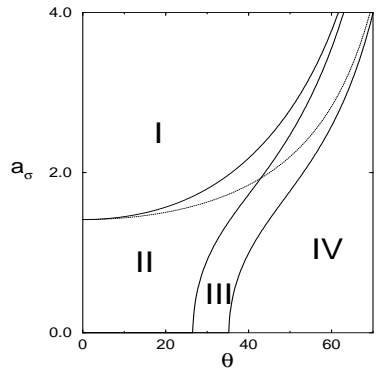,height=2.0in,width=2.0 in}
\caption{Phase diagram for the isotropic case $\sigma=\mu=1$ at $T <
T_c$.  Region I: $\nu_x<0$, $\nu_y < 0$, $D_x^I >0$, $D_y^I > 0$ and
$\ell_x > \ell_y$; Region II: $\nu_x<0$, $\nu_y < 0$, $D_x^I >0$,
$D_y^I > 0$, and $\ell_x < \ell_y$; Region III: $\nu_x<0$, $\nu_y <
0$, $D_x^I <0$ and $D_y^I > 0$; Region IV: $\nu_x>0$, $\nu_y < 0$,
$D_x^I <0$ and $D_y^I > 0$. Note that the phase diagram is independent
of the precise values of $J$ and $p$, while the $\epsilon$ dependence
is contained in $a_\sigma$. }
\label{fig4}
\end{figure}

{\it Morphological phase diagram---} The detailed morphological phase
diagram is rather complex if the diffusion is not thermally activated,
but ion-induced. At low temperatures, when $D_T$ is negligible, the
ripple wavelengths are $\ell_x^I = 2 \pi \sqrt{D^I_x/|\nu_x|}$ and
$\ell_y^I = 2 \pi \sqrt{D^I_y/|\nu_y|}$. In the following we discuss
the expected surface morphologies in function of the experimental
parameters $\theta$ and $a_\sigma$, based on the phase diagram shown
in Fig.  \ref{fig4}.

  {\it Region I---} The surface tensions, $\nu_x$ and $\nu_y$, are
negative, while $D_x$ and $D_y$ are positive, consequently we have a
superimposed ripple structure along the $x$ and the $y$ directions.
The ripple wavelength observed experimentally is the smallest of the
two, and since $\ell^I_x > \ell^I_y$ the ripple wave vector is
oriented along the $y$ direction.  The lower boundary of this region
separating it from Region II is given by the solution of the
$\ell^I_x=\ell^I_y$ equation.

{\it Region II---} Here the ripple wave vector is oriented along the
$x$ direction, since $\ell^I_x < \ell^I_y$. This region is bounded
below by the $D^I_{x}=0$ line.  At large length scales in I and II one
expects kinetic roughening described by the KPZ equation
\cite{alb,revrough1,us,kpz,revrough}.

{\it Region III ---} In this region $D^I_{x}$ is negative, while the
sign of all other coefficients are as in I and II.  Since both the
surface tension and the surface diffusion are destabilizing along $x$,
every mode is unstable and one expects that the KPZ nonlinearity
cannot turn on the KS stabilization \cite{us,kent}, the system being
unstable at large length scales as well, leading to exponential
growth. The lower boundary of this region is given by the
$\nu_x=0$ line.

{\it Region IV ---} Here we have $\nu_x > 0$, $\nu_y<0$, $D_{x}^I <
0$ and $D_{y}^I > 0$, i.e. one expects the surface to be periodically
modulated in the $y$ direction. In the $x$ direction we have an
interesting reversal of the instability: the short length scale
instability generated by the negative $D^I_x$ is stabilized by the
positive surface tension $\nu_x$. Thus there is no ripple structure
along the $x$ direction. Regarding the large length scale behavior,
along the $x$ direction the surface diffusion term is irrelevant
compared to the surface tension, thus one expects KPZ scaling.
However, along the $y$ direction the KS mechanism is expected to act,
renormalizing the negative $\nu_y$ to positive values for length
scales larger than $\ell^I_y$, leading to a large wavelength KPZ
behavior.

If thermal and ion-induced diffusion coexist, the ripple wavelengths
are given by $\ell_x=2 \pi [(D_T + D_{x}^I)/|\nu_x|]^{1/2}$ and
$\ell_y=2 \pi [(D_T + D_{y}^I)/|\nu_y|]^{1/2}$. The phase diagram for
intermediate temperatures can be calculated using the total $D$.  In
particular for high $T$, when $D_T >> D_{x}^I$ and $D_T >> D_{y}^I$,
the phase diagram converges to the one obtained in Ref. \cite{us}, the
ripple orientation being controlled by the $\nu_x=\nu_y$ line (dotted
line in Fig. \ref{fig3}).  Thus with increasing $D_T$ the phase
boundary between the regions I and II converges to the $\nu_x=\nu_y$
line and the $D_{x}=0$ boundary separating the regions II and III
shifts downwards, eventually disappearing. However, in the
intermediate regions new phases with coarsening ripple domains
\cite{rost} appear as the $D^I_{x}=0$ line crosses the $\nu_x=0$ line.

While we limited our discussion to the effect of the ion-induced
diffusion on the surface morphology, the results (\ref{coefs2})
can be used to investigate other phenomena as well, such as island
nucleation. The experimental verification of the above results would
constitute an important step to elucidate the mechanism responsible
for ion-induced diffusion, with potential applications to ion-enhanced
epitaxy as well.

\section{Atomistic models}
\label{M1}

The methods employed for the modeling of growth phenomena at the 
atomic level range from first-principle calculations,  to molecular dynamics, 
and Monte Carlo (MC) simulations \cite{kaxiras}.  
However, currently only MC methods can
reach the long time scales and fairly large length scales needed to
observe ripple formation and roughening.  Moreover, the general
features of ripple formation and roughening are rather generic,
suggesting the existence of a {\it material independent robust
mechanism} governing them.  Thus simple models that incorporate the
basic physics of the system, i.e.  ion-bombardment and surface
diffusion, should be successful in capturing the observed behavior,
see {\em e.g.} Ref.\ \cite{rodo}.
We have developed atomistic MC models of erosion that include
ion-bombardment, ion-induced atom removal and activated surface
diffusion.  To model {\it surface diffusion} we use the methods
developed for MBE, taking the diffusion rate of an atom proportional
to $\exp[-E_a/k_B T]$, where $E_a$ is the activation energy for
diffusion and $T$ is temperature \cite{alb}.
	
To model the ion-bombardment we assume that the ion beam has a
constant flux, but the time and the position when and where an ion
strikes the surface is random.  The ion penetrates the bulk reaching a
penetrating depth $a$ and releases its kinetic energy (see inset of
Fig. 1).  The energy reaching the surface atoms, $E_{ion}$, is
calculated using the energy distribution (5).  If $E_{ion}$ is larger
than the desorption energy $\tilde E_d=n E_B + E_0+E_d$, where $n$ is
the number of nearest neighbors, then it will be sputtered.  If
$E_{ion}$ is smaller than $\tilde E_d$, then it will contribute either
to surface diffusion with probability $\exp[-(n E_B +
E_0-E_{ion})/k_{B}T]$ or to breaking the bonds (sputtering) with
probability $\exp[-(\tilde E_d-E_{ion})/k_{B}T]$.  In the simulation
we used $\epsilon$=1000 eV, $\sigma$=$\mu$=$a$=10.9 lattice constants,
$E_{0}=0.4$ eV, $E_{B}=0.1$ eV, and $E_{d}$=1.0 eV.

	We used the structure factor, $S(k)\equiv<{h}(k,t){h}(-k,t)>$,
where ${h}(k,t)$ is the Fourier transform of $h(x,t)$,  to estimate the
ripple wavelength $\ell$. Typically $S(k)$ develops a 
sharp maximum, which allows us
to estimate $\ell$. However, at large erosion times the structure
factor reflects the roughening of the substrate as well, thus the maximum 
will slowly disappear. Similarly, the maximum is more visible at low
temperatures than at large $T$.

\begin{figure}[bht]
\vskip 0.5 cm
\hskip 3.5 cm
\psfig{figure=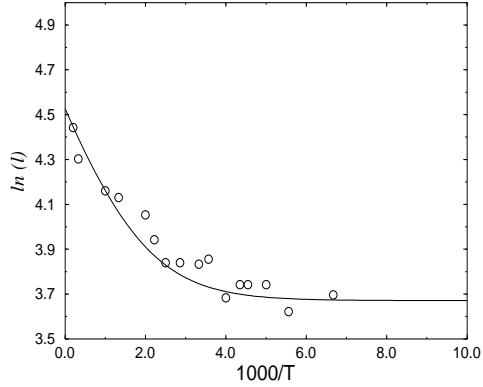,height=2.0in,width=2.5in}
\caption{Ripple wavelength as a function of the inverse temperature
obtained in the numerical simulations of a one dimensional sputtering
model. The continuous line represents an analytical fit (see the
text).}
\label{fig5}
\end{figure}

Fig. \ref{fig5} shows the ripple wavelength as a function of temperature
for $\theta=0$. At large temperatures $\ell$ follows an Arrhenius law,
however, at low $T$ it converges to a constant value. These results
are reminiscent of the experimental data of Maclaren {\it et al.}
\cite{maclaren}, providing direct numerical evidence of ion enhanced
surface diffusion. Indeed, using $\ell = 2 \pi \sqrt{D_T+D^I}$, where
$D_T$ follows an Arrhenius law $D_T = D_0 \exp(-E_a/kBT)$, we can
obtain an excellent fit to the data in Fig. \ref{fig5}. Furthermore,
the fit also allows us to determine the diffusion coefficients in our
simulation, providing $D_0=7000$, $E_a=0.08$eV, and $D^I=1543$. There
are two observations we have to make analyzing these results. First,
$E_a$ is much lower than the activation energy we used as an input to
the simulations. However, this is not in contradiction: the effective
activation energy felt by the diffusing atoms is lowered by the energy
provided by the ions. This effect lowers $E_a$. Indeed, we measured
the average activation energy $E_a^{eff}=E_a-E_i$ for the atoms, the
result agreeing in order of magnitude with the $E_a^{eff}$ obtained
from the fit in Fig. \ref{fig5}. Second, $D_0$ is smaller than the
experimentally expected values, that are of order $10^{xx}$, but 
agrees with the smaller diffusive activity we can obtain in the
numerical simulations (due to running time limitations). It is easy to
see that a larger $D_0$ would lead to a more extended high-temperature
region, such as the one observed by MacLaren {\it et al.}
\cite{maclaren}.

\section{Acknowledgment}

We would like to acknowledge discussions and comments by L.\ A.\ N.\
Amaral, G.\ Grinstein, K.\ B.\ Lauritsen, H.\ Makse, L.\ M.\ Sander,
and H.\ E.\ Stanley.  This research was partially supported by the
University of Notre Dame Faculty Research Program.



\begin{thebibliography}{99}
\bibitem{notexi} In the expansion of (\ref{hv}) two additional
nonlinearities are obtained on the rhs of Eqn.\ (\ref{eqn}):
$\xi_x
\frac{\partial h}{\partial x} \frac{\partial^2 h}{\partial x^2} +
\xi_y \frac{\partial h}{\partial x} \frac{\partial^2 h}{\partial
y^2}$, where $\xi_x$ and $\xi_y$ are functions of $a$, $\sigma$,
$\mu$
and $\theta$ \cite{cb}. One can see that $\xi_x$ and $\xi_y$ are
irrelevant at large length scales using the known values of
$\alpha$
and $z$ \cite{numer} for both the
nonlinear and the linear fixed
points in 2+1 dimensions.
\bibitem{note2}
The terms $-v_0$ and $\partial_x h$ can be reabsorbed by
a change of variables to a comoving frame, and do not
affect the scaling properties.
\end{thebibliography}
\end{document}